
\input Jnl.TEX
\input Reforder.TEX

\def\alp{\alpha}

\def\inf{\infty}

\title {The Dynamics of the Bean Critical State.}

\author W. Barford, W. H. Beere and M. Steer
\affil {\rm Department of Physics, The University of Sheffield,
Sheffield, S3 7RH,
United Kingdom.}

\abstract

A simple  one
dimensional model to simulate the establishment of the Bean critical
state is introduced. It is shown that the
dynamics of the flux lines as they enter the superconductor  are
dominated by `avalanches'. The
distribution of distances moved by vortices in the avalanches
obeys a power law with an exponent of -1. This
suggests that the Bean state is a self-organised critical state.
The density of flux lines is parabolic.


\endtopmatter

\vfill\eject

The motion of flux lines in a type II superconductor
results in phase slip and hence dissipation, since  a potential
difference is required to maintain the phase  difference. If a
current is passed through a superconductor a non-uniform
distribution of flux density will be established. To prevent the
flux lines from sliding, and hence causing dissipation, the flux
lines must be pinned by crystalline defects or random
inhomogeneities. If, however, the magnetic pressure gradient exceeds
the pinning force movement of flux lines arises\refto{degennes}.
This defines the   critical current density, j$_c$. An understanding
of the onset of flux motion is therefore important in understanding
the I-V characteristics of type II superconductors. This is turn
depends on an understanding of the non-equilibrium distribution of
flux line density, and the manner in which it is established.

	The Bean critical state describes the meta-stable distribution of
flux  density in a dirty superconductor\refto{bean}. Observation of
1/f noise when this state is perturbed by external magnetic
fields\refto{kim, ooijen} leads to the suggestion that this critical
state  is in fact a self-organised critical (SOC) state, that is one
which is established dynamically in a  dissipative system and which
is  always on the brink of an instability\refto{bak}. A SOC state
is also characterised by having no intrinsic length and time scales.

To investigate the dynamics of the Bean critical state we have
devised a simple one-dimensional simulation. In this simulation we
have in mind the movement of flux lines into a semi-infinite
superconductor as an external magnetic field is turned on. We
envisage flux lines, which are nucleated on the surface of the
superconductor by the surface current, being driven into the
superconductor by their mutual repulsive interactions. A random
distribution of pinning centres pins the vortices as they
propogate through the superconductor, hence establishing the Bean
critical state.

Let us now describe the model in more detail. Rather than using the
correct inter-vortex potential, namely $K_0(r_{ij}/\lambda)$, we
adopt a simplified short range version introduced by Pla and
Nori\refto{pla} in the interests of numerical simplicity. In
particular, the potential experienced by a vortex at $x_i$ from the
other vortices is, $$
V_{vi}=\sum_{j \ne i} A_v\left(|x_i-x_j|-\xi_v\right)^2, ~{\rm for}
{}~ |x_i-x_j| \leq \xi_v.
\eqno(1)
$$
The mutual force between the $ith.$ and $jth.$ vortex, $f_{ij}=
-dV_i/dx_{ij}$, therefore has a range $\xi_v$ and decreases linearly
with separation.

The distribution of pinning centres, $\{x_\alpha \}$, is taken to
be pseudo-random, that is the pinning centre interactions do not
overlap and there is a maximum permissible separation distance. We
assumed two types of interaction between the pinning centres and the
vortices: a finite range Hook's law,
 $$ V_{pi}=\sum_{\alp} - {\xi_p
\over 2} + {1 \over 2\xi_p}\left(x_i-x_{\alp}\right)^2, ~{\rm for}~
|x_i-x_{\alp}| \leq \xi_p, \eqno(2a) $$
and a sinusoidal dependence,
$$
V_{pi}=\sum_{\alp} - {\xi_p \over \pi}
\left[ 1 + \cos\left(\pi(x_i-x_{\alp})/\xi_p\right) \right],
{}~{\rm for}~ |x_i-x_{\alp}|   \leq \xi_p.
\eqno(2b)
$$
At the surface of the superconductor, $x=0$, there is a force, F,
 of range $\xi_v$ driving in the vortices.

It is convenient to introduce a dimensionless interaction strength,
$\eta$, as the ratio of the work done in removing a vortex from a
pinning centre to the work done in bringing two vortices together,
namely,
$$
\eta = { \int_0^\inf f_{i\alp}(x_{i\alp})dx_{i\alp} \over
\int_\inf^0 f_{ij}(x_{ij})dx_{ij}}.
\eqno(3)
$$
For the vortex-pinning centre interaction (2a) this is,
$
{\xi_p/2 \over A_v\xi_v/2},
$
while for the interaction (2b) it is,
$
{\xi_p/ \pi \over A_v\xi_v/2}.
$
A value of $\eta <<1$, $\sim 1$, and $>>1$ represents weak,
intermediate to strong and very strong pinning, respectively. We
work in the intermediate to strong pinning regime, scale all lengths
by $\xi_v$ and take $\xi_p=\xi_v$.

Let us now describe the simulation. The driving force, F, is turned
on slowly from zero in increments of 0.01. Initially there are no
vortices in the superconductor. At each increment of F vortices are
allowed to enter the superconductor and attain their equilibrium
positions. This is determined by the nett force on each vortex,
resulting from the other vortices and the pinning centres, being
zero. As the vortices enter the superconductor they shunt
other vortices further along, thereby building up a
distribution of flux line density.\refto{footnote1}. By the nature
of this one dimensional simulation each vortex usually has (at
least) two neighbours with which it is interacting. Hence, if a
vortex enters the superconductor all the vortices will move forward.
Typically there is `stick-slip' behaviour: as F is increased
vortices will be unable to enter the superconductor owing to the
outward pressure from the pinned vortices. However, for a
critical value of F this outward force will be overcome and vortices
will flood in causing an `avalanche' of flux motion. It is important
to note that these `avalanches' usually cause all the vortices to
move. Hence, there is no distribution of avalanche sizes as one
expects in a two or three dimensional simulation. However,
there is a distribution of the distances moved by the vortices which
shows scaling behaviour\refto{footnote2}. In figure (1a) we plot the
distribution of vortices moving a distance d for the interaction
(2a) with $\eta=1$. To obtain this plot the force has been increased
from $0$ to $3$ in steps of 0.01. For each increment of F the number
of vortices moving a distance $d \to d+\Delta d$ is recorded. The
figure represents the integrated distribution up to $F=3$, at which
point 192 vortices have entered the superconductor. The slope is
$-1.00$ over three orders of magnitude. Very similar results were
found for the interaction (2b), with $\eta = 2/\pi$, shown in figure
(1b). In this case F was increased from 0 to 1 in increments of
0.001. In the best linear region, from $d=0.001$ to $0.1$, the slope
is close to -1.

We next consider the density of flux lines after 192 have
entered the superconductor. This is shown in figure (2) where the
squares represent the density of vortices at a pinning centre
averaged over neighbouring pinning centres. The dashed line is a
parabolic fit to the data. It is revealing to note that this
density distribution is precisely what one would predict in the
Bean critical state if the average pinning force is independent of
local flux density. To see this we recall that the Bean critical
state is specified when the force from the pinning centres balances
the magnetic force arising from the gradient in the magnetic
pressure\refto{degennes}, namely,
$$
\left| {B \over 4\pi} {dB \over dx} \right| = f_c,
$$
in one dimension.
If $f_c$ is assummed constant, this is easily integrated to
give a magnetic profile of,
$$
B(x) = B(0)\left(1 - {x \over \Lambda} \right)^{1/2},
$$
where  $\Lambda = 8\pi f_cB(0)^2$.

In conclusion, we have introduced a simple  one
dimensional model to simulate the establishment of the Bean critical
state. We showed that the
dynamics of the flux lines as they enter the superconductor  are
dominated by `avalanches'. The
distribution of distances moved by vortices in the avalanches
follows a power law behaviour with an exponent of -1. This would
suggest that the density distribution of flux lines is a
self-organised critical state.  Finally, we found that the density of
flux lines follows a parabolic behaviour, valid in the Bean
model with a constant pinning force.

There are various ways in which this model is deficient. Most
obviously, it is a one-dimensional simulation, so the flux lines
always have two neighbours and disturbances propogate through the
entire system. It also means that flux lines cannot slide pass one
another. Secondly, the correct long range
inter-vortex potential was not used. Such a long range potential
means that in practice each vortex interacts {\it directly} with many
other vortices, leading to the concept of the vortex
bundle\refto{anderson}. This may effect the validity of the
self-organised criticality picture.

Acknowledgements. We thank J E Bishop and G A Gehring for
useful discussions. W. B. also acknowledges support from the SERC
(United Kingdom) (grant ref. GR/F75445).

\vfill
\eject

\head{References}

\refis{degennes} P G de Gennes, Superconductivity of Metals and
Alloys (W A Benjamin, New York, 1966)

\refis{pla} O Pla and F Nori, Phys. Rev. Lett., 67, 919 (1991)

\refis{bean} C P Bean, Phys. Rev. Lett., 8, 250 (1962).

\refis{ooijen} D J Van Ooijen and C P Bean, Rev. Mod. Phys.,
36, 48 (1964).

\refis{kim} Y B Kim, C F Hempstead and A R Strnad, Phys. Rev. B,
131, 2486 (1963)

\refis{bak} P Bak, C Tang and K Wiesenfeld, Phys. Rev. A,
38, 364 (1987).

\refis{anderson} P W Anderson and Y B Kim, Rev. Mod. Phys., 36, 39
(1964)

\refis{footnote1} $\it N.B.$ We do not add an explicit Lorentz
force as is customary with some authors. The force on the vortices
from a current is implicitly included by virtue of the inter-vortex
interactions.

\refis{footnote2} The cellular automata model of a one dimensional
sand pile, introduced by Bak {\it et al.}\refto{bak}, exhibits no
non-trivial scaling laws. Our model is therefore a departure from
this.

\endreferences

\vfill\eject

\head{Figure Captions}

Figure (1a)
The number of vortices (arbitrarily normalised) moving a distance d
versus d for the interaction (2a). The distance is scaled by
$\xi_v$. The external force, F, has been increased from 0 to 3 in
increments of 0.01, with a pinning strength, $\eta = 1$.

Figure (1b)
The same as (1a) with the
interaction (2b). F has been increased from 0 to 1 in increments of
0.001 and $\eta = 2/\pi$.

Figure 2.
The flux density for 192 vortices using the interaction (2a). The
squares represent the average density at a pinning centre, averaged
over neighbouring pinning centres. The dashed line is a parabolic
fit to the data.

\endit \bye